\documentclass[a4paper,11pt]{article}
\pdfoutput=1 % if your are submitting a pdflatex (i.e. if you have
\usepackage{jinstpub} % for details on the use of the package, please
\usepackage{parskip}
\usepackage{float}
\usepackage{bold-extra}
\usepackage{amsmath}
\usepackage{amssymb}
\usepackage{subfigure}
\usepackage{times, txfonts}
\usepackage{mathtools}

\title{\boldmath The CMS RPC Detector Performance and Stability during LHC RUN-$2$}

\author[y,1]{M A Shah,\note{Corresponding author.}}
\author[c,1]{R Hadjiska,}

\author[a]{A. Fagot} 
\author[a]{, M. Gul} 
\author[a]{, C. Roskas} 
\author[a]{, M. Tytgat}
\author[a]{, N. Zaganidis}
\author[b]{, S. Fonseca De Souza}
\author[b]{, A. Santoro}
\author[b]{, F. Torres Da Silva De Araujo}
% ...

\affiliation[a]{Ghent University, Dept. of Physics and Astronomy, Proeftuinstraat 86, B-9000 Ghent, Belgium}
\affiliation[b]{ Dep. de Fisica Nuclear e Altas Energias, Instituto de Fisica, Universidade do Estado do Rio de Janeiro, Rua Sao Francisco Xavier, 524, BR - Rio de Janeiro 20559-900, RJ, Brazil}
\affiliation[c]{Bulgarian Academy of Sciences, Inst. for Nucl. Res. and Nucl. Energy, Tzarigradsko shaussee Boulevard 72, BG-1784 Sofia, Bulgaria.}
% ...
\affiliation[y]{National Centre for Physics, Quaid-i-Azam University, Islamabad, Pakistan.}
% ...

\emailAdd{mashah@cern.ch, roumyana.mileva.hadjiiska@cern.ch}

\abstract{The CMS experiment, located at the Large Hadron Collider (LHC) in CERN, has a redundant muon system composed by three different gaseous detector technologies: Cathode Strip Chambers (in the forward regions), Drift Tubes (in the central region), and Resistive Plate Chambers (both its central and forward regions). All three are used for muon reconstruction and triggering. The CMS RPC system confers robustness and redundancy to the muon trigger. The RPC system operation in the challenging background and pileup conditions of the LHC environment is presented. The RPC system provides information to all muon track finders and thus contributing to both muon trigger and reconstruction. The summary of the detector performance results obtained with proton-proton collision at $\sqrt{s}$ =$ 13$ TeV during 2016 and 2017 data taking have been presented. The stability of the system is presented in terms of efficiency and cluster size vs time and increasing instantaneous luminosity. Data-driven predictions about the expected performance during High Luminosity LHC (HL-LHC) stage have been reported.}


\collaboration[c]{on behalf of the CMS collaboration}
\proceeding{14$^{\text{th}}$ Workshop on Resistive Plate Chambers and related detectors (RPC2018) 
\\
  Puerto Vallarta-Mexico\\
  19-23 February 2018}
\usepackage{float}%
\usepackage{graphicx, subfigure}
\begin{document}
\maketitle
\flushbottom
\section{Introduction}
\label{sec:intro}

One of the key features of the CMS (Compact Muon Solenoid) experiment \cite{b} is its extensive muon system \cite{d}. As a powerful handle to the signature of interesting events, the trigger and reconstruction capabilities for muons are very important. The muon system allows to identify muons produced in many standard model processes, like top quark, W and Z decay, Higgs boson studies and searches beyond the Standard model, as well. Hence a robust and redundant spectrometer is needed to provide efficient muon reconstruction and identification. The CMS muon system exploits three different gaseous technologies, namely, Drift Tubes (DT) in the barrel (central) region, Cathode Strip Chambers (CSC) in the endcap (forward) region, and Resistive Plate Chambers (RPC) \cite{c} in both the barrel and endcap, covering up a pseudo-rapidity region of $|\eta| < 2.4$, where RPCs are installed up to $|\eta| < 1.9$. The muon system has the key functions of muon triggering, transverse momentum measurement, muon identification and charge determination.

During RUN-$2$, to the end of $2017$, data in amount of 86.6 $fb^{-1}$ have been recorded by the CMS detector and RPC system has contributed very efficiently in data taking during the entire period.

\section{CMS RPC Operation and Performance During RUN-$2$}
%%% gas cocentration   %%%%%
The CMS RPCs are used mainly as trigger detectors. Their fast response is guaranteed by the chosen design. They are $1056$ double gap chambers in the RPC system with Bakelite plates with a bulk resistivity in the range of $10^{10} - 10^{11}\, \Omega\cdot$cm. The chambers are working in an avalanche mode which allows them to operate in a high rate of ionizing particles reaching levels more than few hundred Hz/cm$^2$. The intrinsic time resolution is $\approx2$ ns \cite{AbTime}. This is much less than the $25$ ns time window provided by the RPC front-end electronic and thus, the assignment of correct time slot of the registered particles is ensured. The performance of RPCs depends on the usage of proper working gas mixture. In order to operate in avalanche mode the CMS RPCs are using a composition  of $3$ gases, such as 95.2\% Freon ($C_2H_2F_4$) in order to enhance an ionization caused by the incident particle, 4.5\%  Isobutane ($iC_4H_{10}$) used as a quencher gas to reduce streamer formation, and 0.3\% $SF_6$ in order to control the background electrons.

\begin{figure}[!htb]
\centering
\includegraphics[width=.65\textwidth]{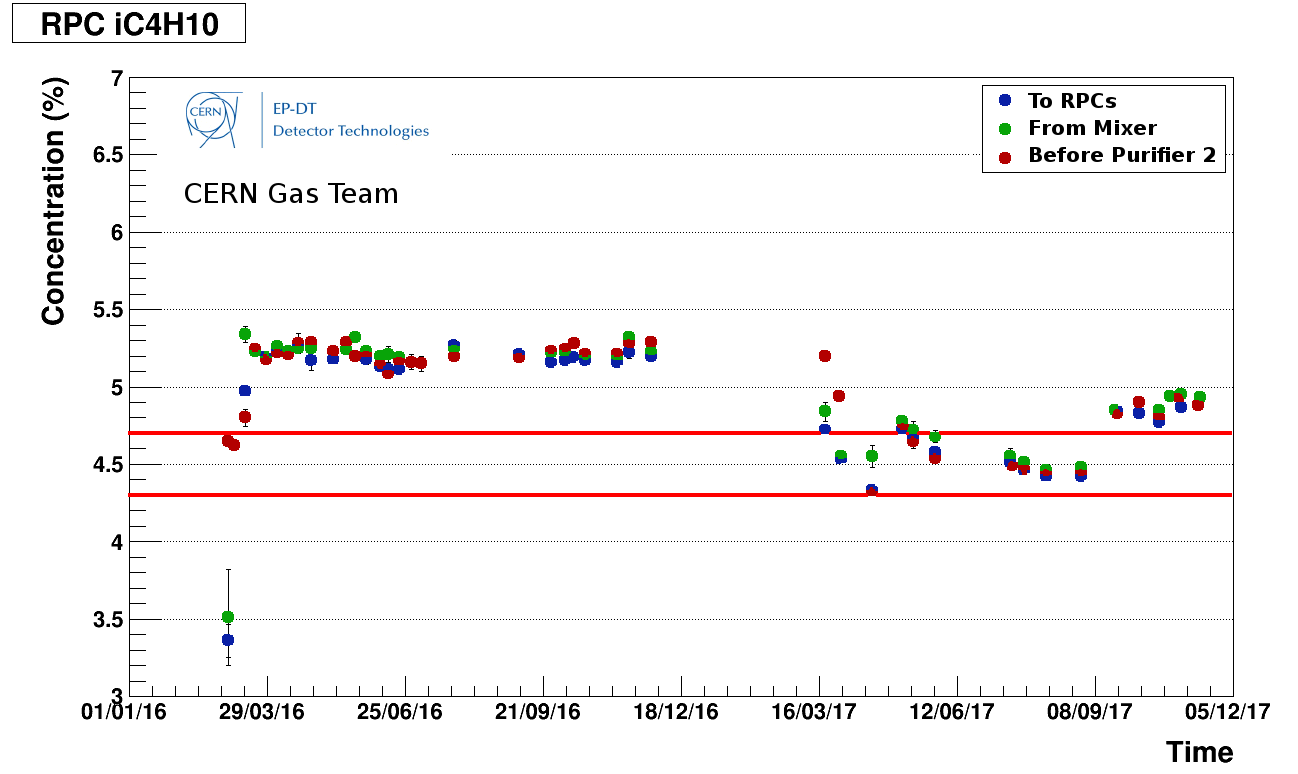}
\caption{Isobutane concentration level in RPC system in 2016 and 2017.}
\label{fig:gg}
\end{figure}

Fig \ref{fig:gg} shows the Isobutane concentration level in the RPC system during 2016 and 2017. The two red lines on the plot mark the limits for optimal performance of the CMS RPCs. With green color is given the Isobutane concentration from the mixer. The gas coming from the system before purifier 2 is shown with red color. The blue dots represent the gas going back to the system. As might be seen from the plot, in 2016 the Isobutane concentration in the RPC gas working mixture was higher. The reason was a problem with the mass flow controller. The effect of the changed Isobutane concentration on the chambers performance will be shown in the next section of this document.

%%%%%%%%
\subsection{CMS RPC Calibration and Dependence on the Gas Mixture Composition}
Important parameters for the RPC system performance monitoring are the RPC hit registration efficiency and cluster size, defined as a number of adjacent strips fired with response to the passage of charged particles. The CMS RPCs are one layer detectors and the reconstructed RPC hits are the information combined with the other trigger primitives based on the segments built in DT or CSC chambers. The RPC hits coordinates are calculated in the gravity center of the formed clusters of fired strips. A larger cluster size can affect the proper estimation of the bending angle of the muon trajectory. In order to follow the muon trigger requirements the cluster size of the RPC hit should be kept not more than $3$ strips. The proper calibration of the detector is based on the analysis of efficiency and cluster size dependences on the applied high voltage.  The HV scan is taken at effective, equidistant voltages in the working range of $[8600, 9800]$ V. The collected data are being analyzed in order to evaluate the optimal high voltage working points (HV WP). More details about the RPC HV scan methodology might be found in \cite{e, f}. The recent results of HV scans during 2017 with comparison to previous years might be found in \cite{rogelio}.

\begin{figure}[!htb]
	\centering
	\includegraphics[width=.65\textwidth]{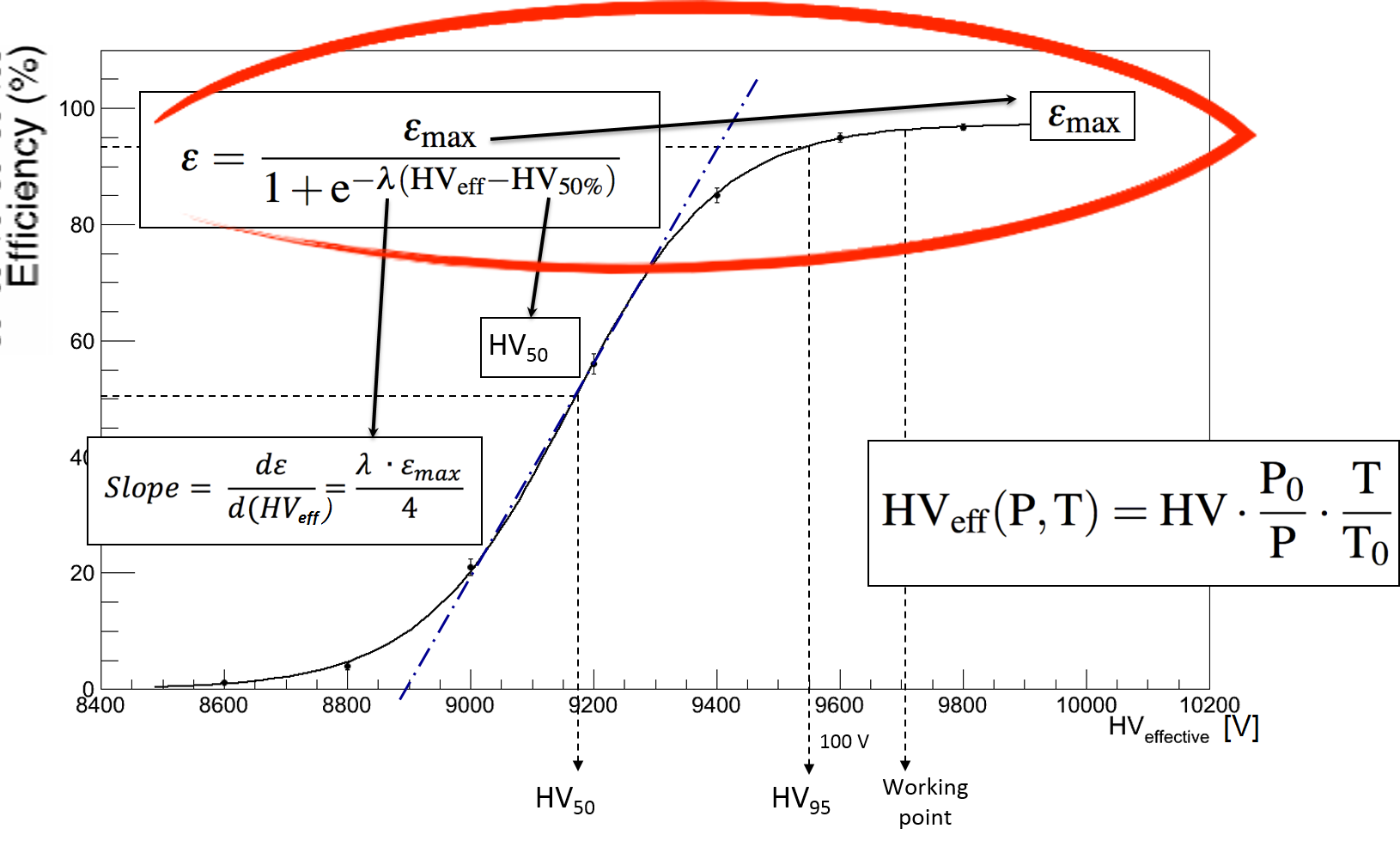}
	\caption{RPC HV Scans - Sigmoid Fit: The efficiency data points, taken at effective voltage (corrected for pressure variations), are fitted by a sigmoid function. For each calibration run the efficiency is calculated for every single RPC eta partition, the smallest RPC granularity object, also known as roll.}
	\label{fig:hvsc}
\end{figure}

The RPC hit efficiency has been calculated with the segment extrapolation method \cite{r3}. The segments built in the nearest DT or CSC chambers have been extrapolated to the plane of the RPC chamber. The segments have been selected to belong to a muon track reconstructed in the muon system. The tracks have been selected to have $p_{T} > 7$ GeV and a quality defined by $\chi^{2}\mathbin{/}ndof < 8$. The coincidences between the extrapolated points and RPC reconstructed hits are searched in a vicinity of $2$ strips. The efficiency has been calculated as the ratio between the number of matched and expected hits. The efficiency data points,
%taken at effective voltage (corrected for pressure variations), are fitted by a sigmoid function \eqref{eq:y:4}.
taken at effective voltage (corrected for pressure variations), are fitted by a sigmoid function as it is shown on Fig. \ref{fig:hvsc}, where $\epsilon_{max}$ is the maximum efficiency, $HV_{eff}$ is effective high voltage, $HV_{50\%}$ is voltage at which the fit efficiency is 50\% of its maximum value and $\lambda$ represents the slope of the sigmoid function and $HV50$.

The HV WP is defined as the voltage at the knee (before the plateau) of the efficiency curve plus 100 V for barrel and 120 V for endcap. The small difference in HV between barrel and endcap detectors depends on few differences in the assembly parameters.

%\begin{equation}
%%\epsilon $=$ \frac{\epsilon_{max}}{1+ \mathrm{e}^{ {-\lambda} \left( HV_{eff} - HV_{50\%} \right)}}
%\epsilon = \frac{\epsilon_{max}}{1 + \mathrm{e}^{{-\lambda} \left(HV_{eff} - HV_{50\%} \right)}}
%\label{eq:y:4}
%\end{equation}

\begin{figure}[!htb]
\centering
\subfigure[]{%
\includegraphics[width=.44\textwidth]{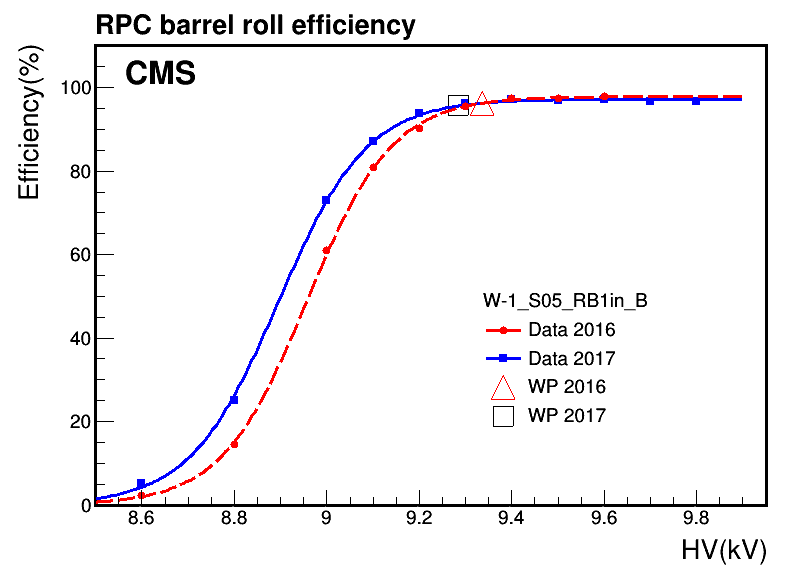}
\label{fig:subfigure1}}
\quad
\subfigure[]{%
\includegraphics[width=.44\textwidth]{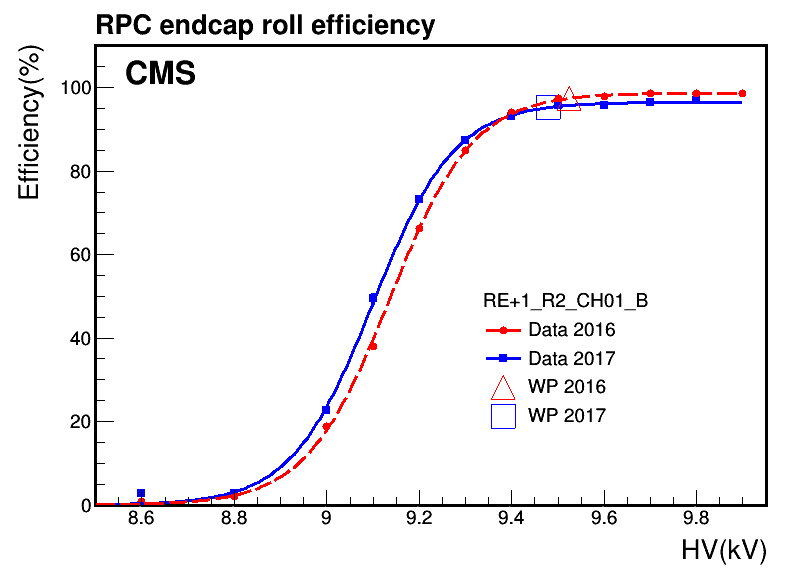}
\label{fig:subfigure2}}
\caption{Efficiency vs HV distributions obtained during 2016 (red) and 2017 (blue) HV scans for two example chambers in barrel (a) and endcap (b).  The shifts of the 2016 curves to higher HV values are caused by the higher Isobutane concentration in 2016.}
\label{fig:effHV}
\end{figure}

The two plots on Fig. \ref{fig:effHV} show the comparison between the efficiency vs HV distributions obtained during 2016 (red) and 2017 (blue) HV scans for two example chambers in barrel and endcap. The shifts of the 2016 (red) curves to higher HV values are caused by the higher Isobutane concentration in 2016. Nevertheless the hit efficiency at working point remains almost unchanged since HV WP both for barrel and endcap (evaluated at 2016 and 2017) are calculated in the plateau of the curves.

\subsection{RPC Efficiency and Cluster Size Stability}

\begin{figure}[!htb]
\centering
\includegraphics[width=.65\textwidth]{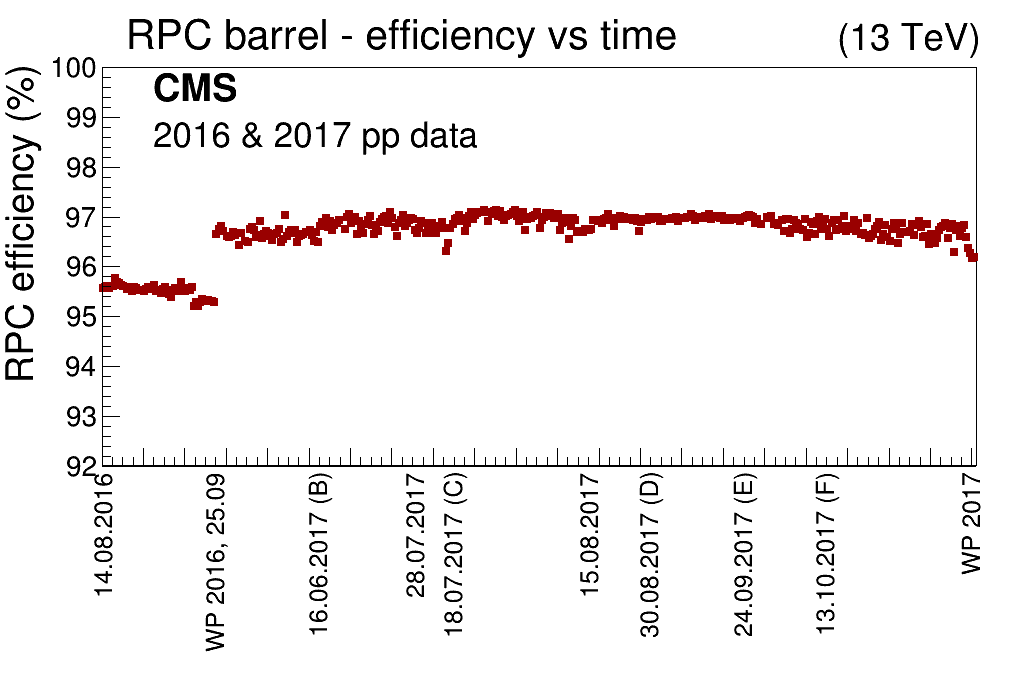}
\caption{Average efficiency of all barrel chambers vs time during 2016 and 2017. The statistical errors for efficiency are less than $0.1\%$ and invisible on the plot.}
\label{fig:1}
\end{figure}

\begin{figure}[!htb]
\centering
\includegraphics[width=.65\textwidth]{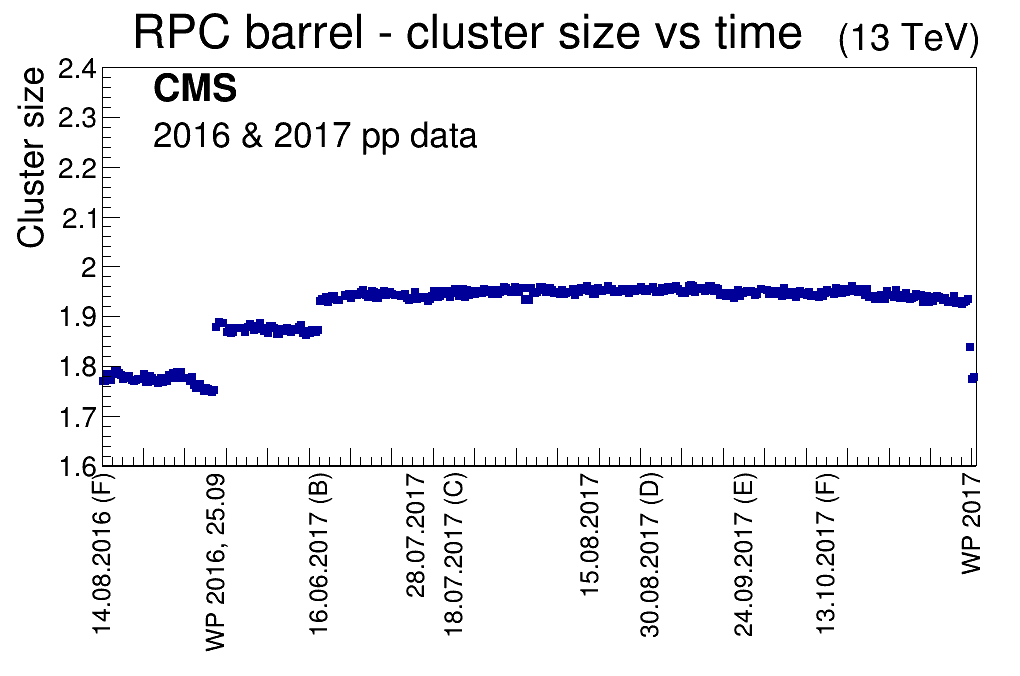}
\caption{Average cluster size of all barrel chambers vs time during 2016 and 2017. The statistical errors for efficiency are less than $0.1\%$ and invisible on the plot.}
\label{fig:2}
\end{figure}

Fig. \ref{fig:1} and Fig. \ref{fig:2} shows the RPC efficiency and cluster size history for the barrel in 2016 and 2017 respectively. Detailed performance results of RPC system for data taken during 2015 can be found in \cite{mehar}. In 2016, because of higher Isobutane concentration (5.3\%), efficiency was lower as the HV working points (WP) were not changed to compensate the wrong gas mixture. After the deployment of the new WP in September 2016, the efficiency increased slightly by ~1\% and cluster size increased sharply. Gas concentration was back at 4.5 \% in 2017 but the WP were not changed. The efficiency remained unchanged (running in the plateau of the sigmoid curve), however a new increase of the cluster size have been observed. New WP have been deployed by end of 2017, which lead to a slight decrease of the efficiency but sensible reduction of the cluster size.

\begin{figure}[!htb]
\centering
\includegraphics[width=.65\textwidth]{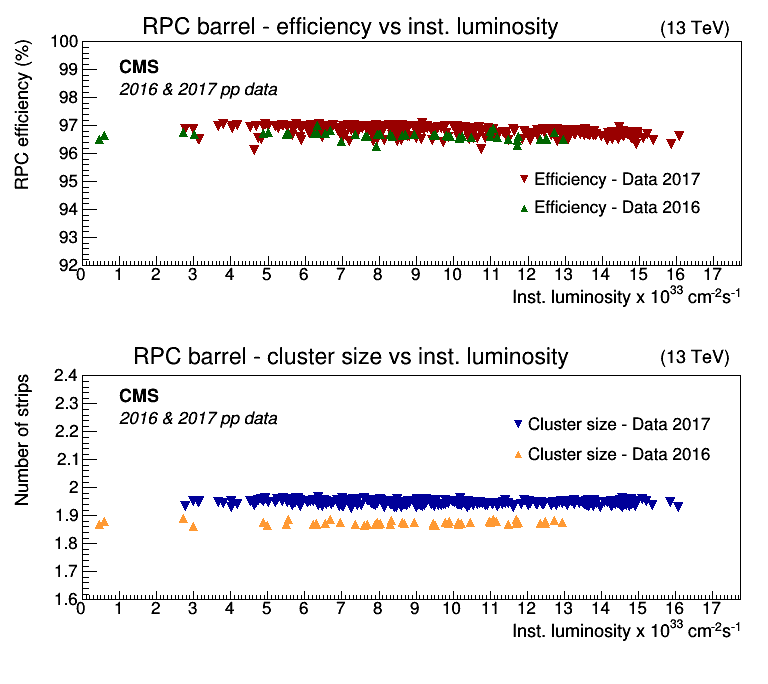}
\caption{RPC barrel efficiency (top) and cluster size (bottom) as a function of the instantaneous luminosity measured in proton-proton collision runs in $2016$ and $2017$ data taking for barrel. The linear extrapolation to instantaneous luminosity of $5\times10^{34}~cm^{-2}s^{-1}$ shows $0.8$\% reduction of efficiency for the barrel.}
\label{fig:3}
\end{figure}

\begin{figure}[!htb]
\centering
\includegraphics[width=.65\textwidth]{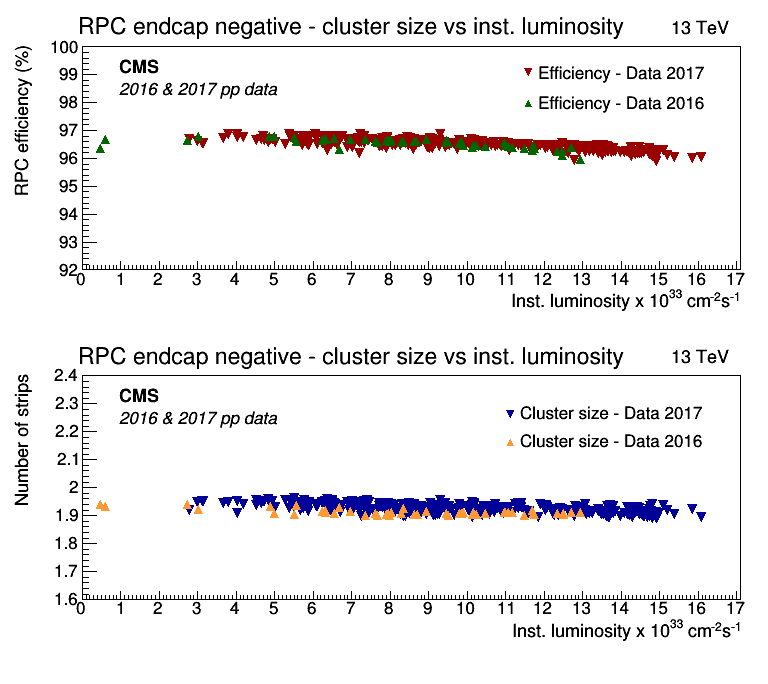}
\caption{RPC barrel efficiency (top) and cluster size (bottom) as a function of the instantaneous luminosity measured in proton-proton collision runs in $2016$ and $2017$ data taking for negative endcap. The linear extrapolation to instantaneous luminosity of $5\times10^{34}~cm^{-2}s^{-1}$ shows $2$\% reduction of efficiency for the endcap.}
\label{fig:4}
\end{figure}

Fig. \ref{fig:3}  and Fig. \ref{fig:4} represents the RPC barrel and negative endcap efficiency and cluster size as a function of the instantaneous luminosity measured in proton-proton collision runs in $2016$ and $2017$ data taking. The data taken at same WP have been used for the comparison. The lower efficiency and cluster size for barrel in $2016$ are caused by the higher Isobutane concentration in $2016$. Nevertheless the comparison between the $2016$ and $2017$ results show stable efficiency and performance. The average cluster size is kept around $2$ and this is far below the maximum limit of $3$ strips, according to the trigger requirements.

Nevertheless the comparison between the $2016$ an $2017$ results show stable efficiency and cluster size. The obtained results were linearly extrapolated to the designed HL-LHC luminosity of $5\times10^{34}~cm^{-2}s^{-1}$ and $0.8$\% reduction of efficiency is found for barrel and $2$\% in the endcap which is consistent with the hit rate in the barrel and endcap as background rate in endcap is twice as that of barrel. No change has been observed for cluster size at HL-LHC conditions from the linear fit applied to the cluster size distributions.

%%%%  Trigger Part %%%%%

\subsection{RPC Trigger Performnace}
With the start of LHC RUN-$2$ in 2015, the energy of the collisions increased from $8$ to $13$ TeV in the center of masses and the instantaneous luminosity reached values larger than $10^{34}~cm^{-2}s^{-1}$. In order to cope with drastically increased total rate, the first level of the trigger system has been upgraded. The logic of the trigger chain of the muon system has been changed, as well. During RUN-$1$ the trigger primitives, formed in the local subsystem triggers were send to the respective track finders and the collected information has been used for the final decision of the global muon trigger. In RUN-$2$ the muon trigger combines the information -- hits and segments - from all muon detectors and send them to three different track finders -- BMTF (Barrel Muon Track Finder), OMTF (Overlap Muon Track Finder) and EMTF (Endcap Muon Track Finder). BMTF uses information from DTs and RPCs and covers pseudo-rapidity region up to $| \eta | \leq 0.83$. In the overlap region, $0.83 \leq | \eta | \leq 1.24$, OMTF combines the information from all the three muon subsystems -- DT, RPC and CSC. In the region above $ | \eta | $ = $ 1.24$, EMTF uses and information from CSC and RPC \cite{r1}. Thus RPC system provides hits to all three track finders. While in the overlap region the RPC hits are sent directly from the link boards to  OMTF, in the barrel they are sent to TwinMux \cite{r2} concentrator card (the adaptive layer for the track finder in the barrel region) and CPPF (Concentrator Pre-Processor and Fan-out ) in the endcap region.

Currently there are three types of trigger primitives seeding the L1 Barrel Muon Track Finder:
	\begin{itemize}
		\item[$\bullet$ ]   DT+RPC segments (in all 4 stations, RPCs are used to complement low quality DT segments);
		\item[$\bullet$ ]  DT-only segments (in all 4 stations, containing only DT information);
		\item[$\bullet$ ]  RPC-only segments (in MB1 and MB2 where two RPC layers are present per station, a linear fit between the inner and outer layer is done to measure the phi direction and the bending of the muon candidate ). The RPC-only segments were enabled to muon trigger in 2017.
	   \end{itemize}

\begin{figure}[!htb]
\centering
\includegraphics[width=.45\textwidth]{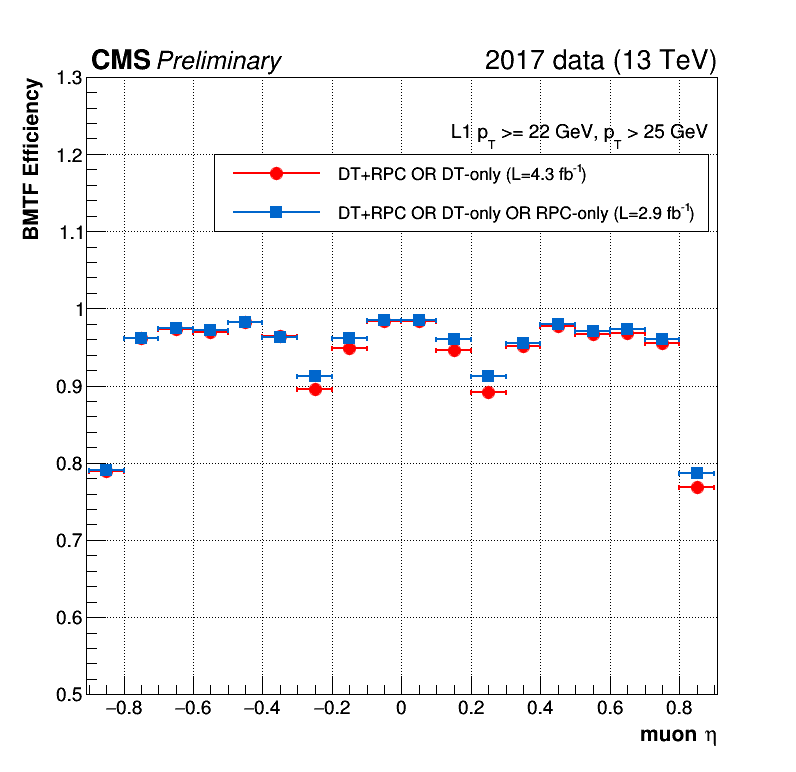}
\caption{BMTF efficiency for muons with $p_{T} > 25~GeV$, with and without inclusion of RPC-only segments in the barrel trigger primitives, as a function of pseudorapidity. The statistical errors are small and almost invisible on the plot. By adding redundancy to the algorithm, up to 2\% higher efficiency is observed in the crack regions (space in between wheels around $| \eta | \approx 0.25~and~ | \eta | \approx 0.85$). The region above $| \eta | $ = $ 0.8$ is also covered by the OMTF.}
\label{fig:trig1}
\end{figure}

\begin{figure}[!htb]
\centering
\includegraphics[width=.45\textwidth]{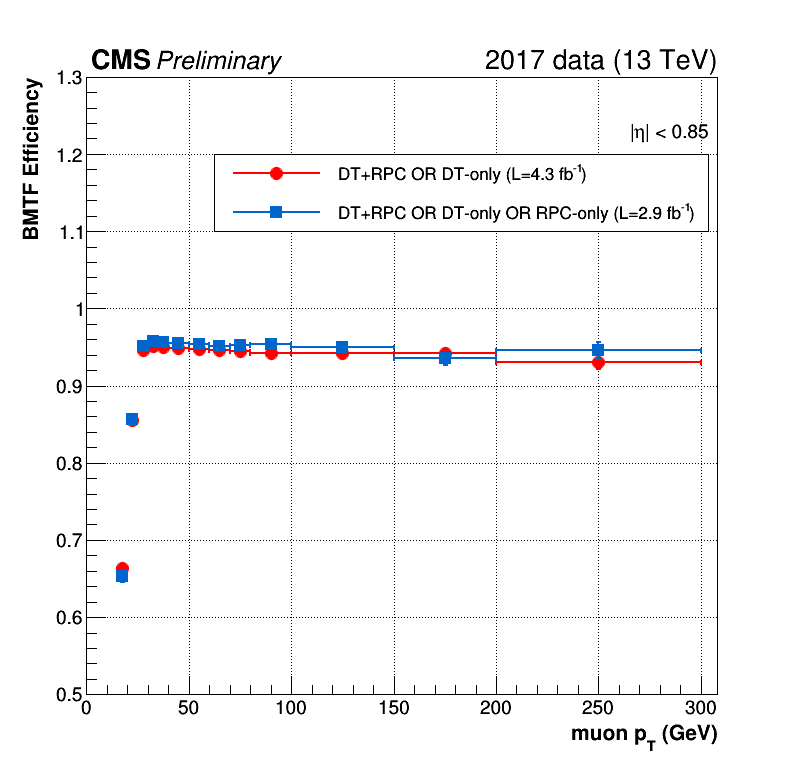}
\caption{BMTF efficiency, with and without inclusion of RPC-only segments in the barrel trigger primitives, as a function of the muon transverse momentum.  The statistical errors are small and almost invisible on the plot. By adding redundancy to the algorithm, the overall BMTF efficiency improves by $\approx 0.7\%$. No degradation in the high $p_{T}$ region is observed.}
\label{fig:trig2}
\end{figure}

The above mentioned primitives can be combined in a logical OR schema. Figures 1 and 2 show the impact of the RPC-only segments on the BMTF performance. The comparison was done for the BMTF efficiency with and without usage of RPC-only segments. The efficiency measurement was done with Tag and Probe method \cite{r3}. Muons with transversal momentum $p_{T} >~25 GeV$, coming from $Z$ decay, have been selected for the analysis, following the identification requirements in \cite{r4}. BMTF efficiency for muons with $p_{T} > 25~GeV$, with and without inclusion of RPC-only segments in the barrel trigger primitives, as a function of pseudorapidity is shown in Fig.~\ref{fig:trig1} By adding redundancy to the algorithm, up to 2\% higher efficiency is observed in the crack regions (space in between wheels around $ | \eta |  \approx 0.25$ and $ | \eta |  \approx 0.85$). The region above $ | \eta | $ = $ 0.8$ is also covered by the OMTF. On the Fig.~\ref{fig:trig2} is presented the BMTF efficiency, with and without inclusion of RPC-only segments in the barrel trigger primitives, as a function of the muon transverse momentum. By adding redundancy to the algorithm, the overall BMTF efficiency improves by $\approx 0.7\%$. No degradation in the high $p_{T}$ region is observed.

%%%% yrigger end  %%%%%%

\section{Conclusion}
CMS RPCs have been operating very successfully during RUN-$2$. The entire system has been included in the new muon trigger logic, contributing to the three muon track finders. After 3 year of LHC running with increasing instantaneous luminosity and several years from the end of RPC construction, the detector performance is within CMS specifications and stable with no degradation observed. No significant issues were found for running up to high luminosity scenarios at HL-LHC.

\acknowledgments We would like to thank especially all our colleagues from the CMS RPC group and L1 muon trigger group for their dedicated work to keep the stable performance of the RPC system. We wish to congratulate our colleagues in the CERN accelerator departments for the excellent performance of the LHC machine. We thank the technical and administrative staff at CERN and all CMS institutes.


\end{document}